\documentclass[setspace,amsmath,amssymb]{article}
\usepackage{setspace}
\usepackage{graphicx}
\usepackage{amsmath}
\usepackage{amssymb}
 
\newcommand{\rrangle}{\rangle\!\rangle}

\newcommand{\ra}{\rangle}
\newcommand{\la}{\langle}
\begin{document}
\begin{titlepage}

\begin{flushright}
\today
\end{flushright}

\vspace{1in}

\begin{center}

{\bf Projection hypothesis in the setting for the quantum Jarzynski equality}

\vspace{1in}

\normalsize

\renewcommand\thefootnote{\fnsymbol{footnote}}

{Eiji Konishi\footnote[0]{E-mail address: konishi.eiji.27c@kyoto-u.jp}
}

\normalsize
\vspace{.5in}

{\it Graduate School of Human and Environmental Studies,\\
 Kyoto University, Kyoto 606-8501, Japan}
\end{center}

\vspace{1in}

\baselineskip=24pt
\begin{abstract}
Projective quantum measurement is a theoretically accepted process in modern quantum mechanics.
However, its projection hypothesis is widely regarded as an experimentally established empirical law.
In this paper, we combine a previous result regarding the realization of a Hamiltonian process of the projection hypothesis in projective quantum measurement, where the complete set of the orbital observables of the center of mass of a macroscopic quantum mechanical system is restricted to a set of mutually commuting classical observables, and a previous result regarding the work required for an event reading (i.e., the informatical process in projective quantum measurement).
Then, a quantum thermodynamic scheme is proposed for experimentally testing these two mutually independent theoretical results of projective quantum measurement simultaneously.
\end{abstract}

\vspace{.1in}


\vspace{.6in}

\end{titlepage}

\setcounter{footnote}{0}

\renewcommand{\thefootnote}{\alph{footnote}}

\section{Introduction}

Projective quantum measurement is a theoretically accepted process in modern quantum mechanics.
For example, it is used in measurement-based feedback processes in the quantum thermodynamics of information \cite{SU1,MT,Nature1} and in standard quantum measurements in quantum computation \cite{NC}.
However, its projection hypothesis is widely regarded as an experimentally established empirical law \cite{SUBook}.

In this paper, we combine two mutually independent theoretical results of projective quantum measurement.
The first, from Ref. \cite{EPL}, regards the realization of a Hamiltonian process of the projection hypothesis in projective quantum measurement, where the complete set of the orbital observables of the center of mass of a macroscopic quantum mechanical system is restricted to a set of mutually commuting classical observables \cite{Neumann,Jauch,dEspagnat}.
The second, from Ref. \cite{JSTAT1}, regards the work required for an event reading (i.e., the informatical process in projective quantum measurement).
We also propose a quantum thermodynamic scheme to experimentally test these two theoretical results simultaneously.

Throughout this paper, to model projective quantum measurement, we adopt the ensemble interpretation of quantum mechanics \cite{dEspagnat}.
Specifically, for a given quantum system, a {\it pure} ensemble means an ensemble of copies of a state vector, and a non-trivial {\it mixed} ensemble means a statistical mixture of copies of at least two distinct state vectors.
Supposing an eigenbasis of the state space (we take it as the eigenbasis of the  discrete measured observable), we refer to an ensemble as a {\it coherent} ensemble when its elements have quantum coherence (i.e., are quantum superpositions of at least two distinct eigenstates) and as a {\it classical} ensemble when its elements do not have any quantum coherence (i.e., are eigenstates).

Projective quantum measurement has two temporally successive processes: {\it non-selective measurement} and its subsequent {\it event reading}.\footnote{Intuitively, the term {\it non-selective measurement} means that, after only this dynamical process, events are in a mixture and no particular event is yet singled out.}
Non-selective measurement (also called a complete {\it decoherence process} or a {\it quantum-to-classical transition} \cite{Zurek}) means a dynamical process from a coherent pure state (that is, the state to be measured) to its classical mixed state (i.e., its diagonal part as a mixture of eigenstates).
Event reading means an informatical process from a classical mixed state to a classical pure state (i.e., an eigenstate).
We call the substance of the latter process the {\it projection hypothesis} (for its older form, see Refs. \cite{Neumann} and \cite{Luders}).

In Ref. \cite{EPL}, the projection hypothesis was derived in a Hamiltonian process after the non-selective measurement of the discrete meter variable of an event reading quantum mechanical system $\psi$ \cite{JSTAT2} by restricting the complete set of the orbital observables of the center of mass of a macroscopic quantum mechanical system to a set of mutually commuting classical observables \cite{Neumann,Jauch,dEspagnat}.
The core idea in this derivation is to use a spatiotemporally inhomogeneous macroscopic Bose--Einstein condensate (BEC) in quantum field theory, where the spatial translational symmetry is broken spontaneously, under the von Neumann-type interaction with the event reading system $\psi$ in the interaction picture.
Due to the Nambu--Goldstone theorem, the center of mass velocity of the condensate is returned to the velocity of a c-number spatial coordinate system in the rearranged spatial coordinate system in the order parameter (i.e., the vacuum expectation value of the boson Heisenberg field) of this condensate \cite{Umezawa1,Umezawa2,Umezawa3,Umezawa4,Umezawa}.
Then, when the macroscopic BEC has a non-relativistic center of mass velocity classically fluctuating in the ensemble, a constraint on the classical mixed state of the event reading system $\psi$ arises after the von Neumann-type interaction between this system $\psi$ and the BEC.
This constraint is the physical equivalence (the Galilean relativity) between {\it state vectors} (equivalent to mixtures of eigenstates of the discrete meter variable with respect to the statistical data of the observables \cite{JSTAT2}) with distinct their relative phases of $\psi$, obtained by the partial trace of the BEC decoupled from $\psi$ in the mixture.
This physical equivalence is realized via the Galilean transformation of the rearranged spatial coordinate system.
This constraint is the trigger of event reading by $\psi$.

In Ref. \cite{JSTAT1}, on the other hand, the projection hypothesis was treated as a {\it black box}.
The novelty of the present work lies in the following two aspects.
First, the derivation of the result in Ref. \cite{JSTAT1} is refined.
Second, a quantum thermodynamic scheme is modeled to experimentally test the result in Ref. \cite{JSTAT1} based on the result of Ref. \cite{EPL}.

Here, we identify the novelty of the present work within the existing literature on quantum thermodynamics.
At present, the energetics of projective quantum measurement is based on the quantum thermodynamics of information \cite{SU1,Nature1,SU2}.
In this theory, the fundamental energy cost is considered for projective quantum measurement, which is done in the general positive operator-valued measure (POVM) measurement \cite{NC}, in the ensemble description \cite{SU1,MT} and the reset process of the measurement outcome stored in the memory system \cite{SU2}.
On the other hand, the present paper and Refs. \cite{JSTAT1} and \cite{JSTAT2} consider the work required for event reading, that is, the informatical process in projective quantum measurement, which is independent from the former energy cost.

Throughout this paper, we denote operators with a hat.

The rest of this paper is organized as follows.
In Sec. 2, we explain the systems to be considered, the treatment of a macroscopic quantum system, and the von Neumann-type interaction.
Section 3 is devoted to a new and clear elaboration on the origin of the required work for event reading derived in Ref. \cite{JSTAT1}.
In Sec. 4, we model the quantum thermodynamic scheme, to experimentally test the previously obtained mutually independent theoretical results of projective quantum measurement simultaneously, and clarify the classical nature of the required work for event reading.
In Sec. 5, we conclude the paper.
In Appendices A and B, we present supplementary calculations.
In Appendix C, we clarify the experimental platform for realizing the systems which appear in the quantum thermodynamic scheme.

\section{Preliminaries}

In this paper, using the notation of Ref. \cite{EPL}\footnote{However, we denote $\Psi$ in Ref. \cite{EPL} by $M$ here.}, we consider four systems in two groups:
\begin{equation}
(S_0\;,\ A^\prime)_S\;,\ \ (\psi\;,\ A)_M\;.\label{eq:6}
\end{equation}
These four systems are defined as follows:
\begin{enumerate}
\item[(i)] $S_0$ is the measured system with a discrete observable $\widehat{{\cal O}}$ to be measured.
\item[(ii)] $A^\prime$ is a macroscopic measurement apparatus.
\item[(iii)] $\psi$ is a quantum mechanical system (the event reading system) with a discrete meter variable $\widehat{{\mathfrak M}}$.
\item[(iv)] $A$ is a spatiotemporally inhomogeneous macroscopic BEC in quantum field theory and thus breaks the spatial translational symmetry spontaneously.
\end{enumerate}
Projective quantum measurement is realized as a controlled redefined Hamiltonian process of the whole system (\ref{eq:6}).
This process is governed by the Schr$\ddot{{\rm o}}$dinger equation in the Schr$\ddot{{\rm o}}$dinger picture under restriction of the complete set of the orbital observables due to the macroscopicity of $A^\prime$ and $A$.
There is no longer any black box.

Here, {\it the Hamiltonian is redefined} in the following sense.
For a {\it macroscopic} quantum mechanical system, whose degrees of freedom are abstracted to the degrees of freedom of its center of mass, we argue that its canonical variables (i.e., the center of mass position $\widehat{Q}^a$ and the total momentum $\widehat{P}^a$) are redefined so that they commute with each other and have simultaneous eigenstates.
By incorporating measurement errors (root-mean-square errors) from the original canonical variables into the simultaneous eigenstates of the redefined canonical variables, these eigenstates form a complete orthonormal system, with discretized simultaneous eigenvalues as Planck cells: this is von Neumann's theorem in Ref. \cite{Neumann}.
Before the redefinition, of course, the canonical commutation relation
\begin{equation}
[\widehat{Q}^a,\widehat{P}^b]=i \hbar \delta_{ab}
\end{equation}
holds.
However, after the redefinition,
\begin{equation}
[\widehat{{\cal Q}}^a,\widehat{{\cal P}}^b]=0\label{eq:SSR}
\end{equation}
holds in the change of notation from $\widehat{Q}^a$ (respectively, $\widehat{P}^a$) to $\widehat{{\cal Q}}^a$ (respectively, $\widehat{{\cal P}}^a$).
Redefining the orbital observables (including the canonical variables and the Hamiltonian) restricts the complete set of the orbital observables to a set of mutually commuting classical observables.
We apply the abstraction and the redefinitions to the macroscopic systems $A^\prime$ and $A$ in Eq. (\ref{eq:6}).
Then, the quantum states of $A^\prime$ and $A$ are, in general, non-trivial classical mixed states of Planck cells due to the absence of coherence between simultaneous eigenstates with respect to the observables \cite{Jauch}.
We call this procedure of the redefinition the {\it orbital superselection rule}.
It plays a crucial role in both the non-selective measurement and event reading.

In quantum mechanics, the orbital superselection rule is required for describing the center of mass of a macroscopic quantum mechanical system when we describe the interaction between a microscopic quantum mechanical system and the center of mass of the macroscopic quantum mechanical system \cite{dEspagnat,Araki1,Ozawa}.
The validity of this procedure is essentially the same as the validity of the coarse-graining of the $\mu$-space distribution of a system, in the thermodynamic limit, in classical statistical mechanics: both procedures reduce a pure state to a mixed state due to the macroscopicity of the system.

Next, for the use in Sec. 4, we explain the von Neumann-type interaction.
A {\it von Neumann-type interaction} between a measured system, $S$, with a discrete measured observable $\widehat{{\cal O}}$ and a measurement apparatus, $A$, with the center of mass position variable $\widehat{Q}^x$ and total momentum variable $\widehat{P}^x$ in a particular spatial dimension $x$ is defined by
\begin{equation}
\widehat{H}_{\rm v. N.}=-\Lambda \widehat{{\cal O}}\otimes \widehat{P}^x\label{eq:HvN}
\end{equation}
for a positive-valued c-number constant $\Lambda$ by the passive means of spatial translation of $A$ in quantum mechanics.\footnote{For accounts for the passive means of spatial translation in quantum mechanics, see Chapter 6 of Ref. \cite{QM}.}
Usually, we assume that $\Lambda$ is strong enough to ignore the free Hamiltonians of $S$ and $A$ relatively to this interaction.

In the presence of this interaction, the center of mass position $Q^x$ of $A$ has an ${\cal O}$-dependent passive spatial translation (a gauge transformation) without inertia; note that the difference between two passive spatial translations is the c-number distance between their resultant reference points and thus is not a gauge.
This is our original argument.
See Fig. 1.

\begin{figure}[htbp]
\begin{center}
\includegraphics[width=0.6 \hsize, bb=0 0 260 204]{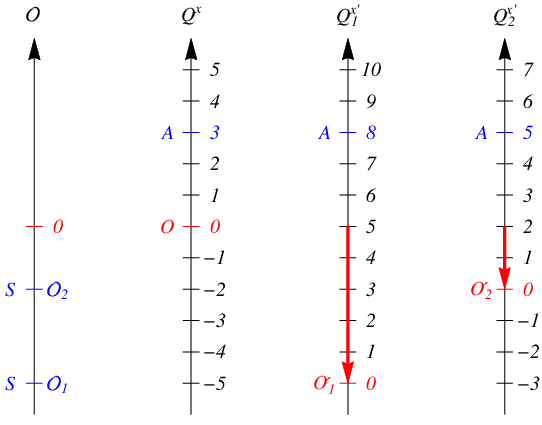}
\end{center}
\caption{Schematic of the von Neumann-type interaction (\ref{eq:HvN}).
Here, we assume two different passive spatial translations from $Q^x$, which result in $\widehat{Q}_1^{x \prime}=\widehat{Q}^x+5$ for $|S\ra=|{\cal O}_1\ra$ and $\widehat{Q}_2^{x \prime}=\widehat{Q}^x+2$ for $|S\ra=|{\cal O}_2\ra$ while keeping $\widehat{P}^x$ constant in the Heisenberg picture of $A$.
The spatial translation of the reference point from $O$ to $O_1^\prime$ and that from $O$ to $O_2^\prime$ are gauge transformations without inertia.
However, the differences $\widehat{Q}_1^{x \prime}-\widehat{Q}^x=5$, $\widehat{Q}_2^{x \prime}-\widehat{Q}^x=2$, and $\widehat{Q}_1^{x \prime}-\widehat{Q}_2^{x \prime}=3$ are c-number distances and thus are not gauges.
In the quantum measurement processes of $\widehat{{\cal O}}$, these differences play a crucial role via the relative phases in the ${\cal O}$-eigenbasis in the presence of the orbital superselection rule of $A$.}
\end{figure}

This interaction was introduced by von Neumann in Ref. \cite{Neumann} to realize the following quantum entanglement between the measured coherent pure state of $S$ in the eigenbasis $\{|{\cal O}\ra_S\}$ of $\widehat{{\cal O}}$ and the ready quantum state $|0\ra_A$ of $A$ in the eigenbasis $\{|Q^x\ra_A\}$ of $\widehat{Q}^x$ after this interaction $\widehat{H}_{\rm v.N.}$ occurs over an elapsed time $\mu$ dominantly:
\begin{equation}
\sum_n c_n|{\cal O}_n\ra_S|0\ra_A\to \sum_nc_n|{\cal O}_n\ra_S |-\mu \Lambda {\cal O}_n\ra_A\;,
\end{equation}
where we adopt the Schr$\ddot{{\rm o}}$dinger picture in the absence of the orbital superselection rule of $A$.
In the state after the interaction, a conventional quantum measurement of $\widehat{{\cal O}}$ is performed by tracing out the state of $A$ \cite{Neumann}.
This measurement is the realization of the classical mixed state of $S$ with respect to ${\cal O}$.

This interaction is also used to realize two processes.
The first is a non-selective measurement of the discrete measured observable $\widehat{{\cal O}}$ of the measured system $S_0$ as a Hamiltonian process in the presence of the orbital superselection rule of a macroscopic measurement apparatus $A^\prime$ \cite{JSTAT1,JSTAT2,Ozawa}.
The second is an event reading process by a quantum mechanical system $\psi$ in a Hamiltonian process in the presence of the orbital superselection rule of a macroscopic BEC $A$ \cite{EPL}.
This second use was explained in the Introduction.

\section{Work Required for Event Reading}

Next, the result of Ref. \cite{JSTAT1} asserts that, for a projective quantum measurement, entropy production of $\sigma=1$ nat (respectively, $-\sigma=-1$ nat) without absorption of heat \cite{JSTAT1} is required in the combined measuring system $M=\psi+A$ (respectively, the combined measured system $S=S_0+A^\prime$) for the event reading (e.r.).
This entropy production is converted to an amount of work required to be done by the combined measuring system $M$ to the combined measured system $S$:
\begin{equation}
W_{\rm e.r.}=k_BT\sigma\;,\ \ \sigma=1\;.\label{eq:0}
\end{equation}
Here, we assume that the combined measured system $S$ is in a non-equilibrium and controlled redefined Hamiltonian process and is initially in thermal equilibrium at temperature $T$ and until then in thermal contact with a heat bath at temperature $T$ \cite{Talkner}.

In the following, we elaborate on the {\it origin} of this entropy production $\sigma$, which accompanies instantaneous relaxation of the system.

Let $\rho=\rho(t)$ denote the statistical weight of the system {\it without} relaxation in the ensemble of copies of the system at a given time $t$.
We set $\rho(0)=1$.\footnote{When we identify $\rho$ with the normalization factor of a density matrix, the relaxation process is truncated from the time evolution of the density matrix.
See Appendix A.1.}

The {\it direct description} in Ref. \cite{JSTAT1} gives rise to the relaxation process in the conventional ensemble at $t=dt$:
\begin{equation}
-\frac{d\rho}{dt}=\delta(t)\;.\label{eq:1}
\end{equation}
The solution to this is $\rho=1-\theta(t)$\footnote{$\theta(t)$ is the Heaviside unit step function that satisfies $d\theta(t)/dt=\delta(t)$ for the Dirac delta function $\delta(t)$.}.
Here, $\delta(t)dt=d\theta(t)$ is the production of $-\rho$.
Namely, in this description, the relaxation occurs in the conventional ensemble at $t=dt$ as
\begin{equation}
-d\rho=1\ \ {\rm at}\ \ t=dt\;.
\end{equation}

The {\it statistical description} in Refs. \cite{JSTAT1} and \cite{GRW} gives rise to a self-similar (i.e., statistical) relaxation process in the enlarged ensemble of the individual conventional ensembles at $t=dt$:
\begin{equation}
-\frac{d\rho}{dt}=\delta(t)\rho\;.\label{eq:2}
\end{equation}
The solution to this is $\rho=\exp(-\theta(t))$.
Here, $\delta(t)dt=d\theta(t)$ is the relaxation {\it probability} in the time interval $dt$ for the individual conventional ensembles in the enlarged ensemble.
Namely, in this description, the self-similar relaxation occurs in the enlarged ensemble at $t=dt$ with unity relaxation probability as
\begin{equation}
-d\rho=\rho\ \ {\rm at}\ \ t=dt\;.\label{eq:fig1}
\end{equation}
This is also obtained by the cutoff of the one-time Poisson process $\rho=\exp(-t/dt)$ with the characteristic time $dt$, which satisfies
\begin{equation}
-d\rho=\rho\ \ {\rm per}\ \ dt\;,\label{eq:fig2}
\end{equation}
at the average occurrence time $t=dt$.

Here, the process (\ref{eq:2}) in the enlarged ensemble is determined by two requirements: {\it self-similarity} with respect to the statistical weight $\rho$, and the {\it unity relaxation probability}; self-similarity means independence of the process equation (\ref{eq:2}) from the value of the statistical weight $\rho$ throughout the process.

For the entropy production $\sigma:=-\ln \rho$ attributable to the decay of $\rho$ from unity, Eq. (\ref{eq:2}) is analogous to a relaxation process in the conventional ensemble at $t=dt$
\begin{equation}
\frac{d\sigma}{dt}=\delta(t)\;,\label{eq:4}
\end{equation}
the solution to which is $\sigma=\theta(t)$.
Here, $\delta(t)dt=d\theta(t)$ is the production of $\sigma$.

In the statistical description, the self-similar relaxation means the {\it entropy production} $\sigma$ of 1 nat with respect to the statistical weight of the system {\it without} relaxation in the enlarged ensemble.
This entropy production is attributable to the fact that, in the statistical description, the relaxation is in a cut-off stochastic process, and thus its occurrence is not mechanical but informative as opposed to the direct description.

In Ref. \cite{JSTAT1}, the instantaneous {\it relaxation} is the temporally contracted {\it non-selective measurement}, and we have elaborated on the origin (\ref{eq:fig1}) of the entropy production $\sigma$ (i.e., the information loss) in Eq. (\ref{eq:0}).

As shown in the next section, this 1 nat of information with respect to the statistical weight of the system {\it without} relaxation in the enlarged ensemble is lost by the combined measuring system $M$ and is acquired by the combined measured system $S$ just before event reading by $M$.

Namely, the entropy production $\sigma$ is partitioned in whole to system $M$.
This partitioning {\it in whole} is because of the time-evolution independence between systems $S$ and $M$ with respect to the absence of the relaxation process \cite{JSTAT1}.
Here, the {\it time-evolution independence} between systems $S$ and $M$ means that any time-dependent process in the composite system of systems $S$ and $M$ is divisible into parts for $S$ and $M$ \cite{JSTAT1}.

\section{Quantum Thermodynamic Scheme}

In this section, we model a quantum thermodynamic scheme to test Eq. (\ref{eq:0}).
Its processes are non-equilibrium and controlled redefined Hamiltonian processes in the setting of the quantum Jarzynski equality \cite{Talkner} with respect to the combined measured system $S$.
These processes consist of five temporally successive steps (apart from the projective energy measurements of system $S$ by the experimenter at the initial and final times):
\begin{enumerate}
\item[(I)] Preparation of the initial quantum state of the whole system (\ref{eq:6}).
In this state, the combined measured system $S$ is in thermal equilibrium at temperature $T$, and the combined measuring system $M$ is in an ${\mathfrak M}$-eigenstate and is decoupled from $S$.

\item[(II)] Preparation of a coherent pure state of $S_0$ to be measured with respect to the discrete observable $\widehat{{\cal O}}$ by using a time-dependent Hamiltonian of $S_0$.

\item[(III)] Non-selective measurement of the discrete observable $\widehat{{\cal O}}$ of $S_0$ by the apparatus $A^\prime$.
For example, this can be realized by means of a dominant and quantum mechanically long-term von Neumann-type interaction between $S_0$ and $A^\prime$ \cite{JSTAT1,JSTAT2,Ozawa,Araki2}.

\item[(IV)] An entangling process of the quantum states of the systems $S$ and $M$ in the eigenbases of $\widehat{{\cal O}}$ and $\widehat{{\mathfrak M}}$, respectively \cite{JSTAT1}.
As a technical requirement, we assume that this entangling process does not change the state of $S$.

\item[(V)] A dominant von Neumann-type interaction between $\psi$ and $A$ to perform the event reading of $\widehat{{\mathfrak M}}$ as mathematically demonstrated in Ref. \cite{EPL}.
\end{enumerate}

Here, we make three remarks on the details of these steps.
\begin{enumerate}
\item[(A)]

We perform the projective energy measurements of $S$ at the initial and final times of this process.
For part $A^\prime$ of $S$, we need the following considerations.
The redefined Hamiltonain $\widehat{{\cal H}}^S$ of $S$ and the redefined canonical variables $\widehat{{\cal Q}}^a$ and $\widehat{{\cal P}}^a$ of $A^\prime$ commute with each other:
\begin{equation}
[\widehat{{\cal H}}^S,\widehat{{\cal Q}}^a]=
[\widehat{{\cal H}}^S,\widehat{{\cal P}}^a]=0\;.
\end{equation}
Then, these three types of operators have simultaneous eigenstates.
This means that the energy eigenstates of $\widehat{{\cal H}}^S$ are individually degenerate with respect to $A^\prime$ for the combinations of ${\cal Q}^a$ and ${\cal P}^a$, each of which gives the energy eigenvalue.
From this fact, the projection operator of energy (i.e., the energy eigenstate obtained by the projective energy measurement of the canonical thermal equilibrium quantum state) with respect to $A^\prime$ is
\begin{equation}
|E\ra\la E|=\sum_{\{({\cal Q}^a,{\cal P}^a)\}_E}|E,{\cal Q}^a,{\cal P}^a\ra\la E,{\cal Q}^a,{\cal P}^a|\;.\label{eq:En}
\end{equation}
Here, we define
\begin{equation}
|E\ra=\bigoplus_{\{({\cal Q}^a,{\cal P}^a)\}_E}|E,{\cal Q}^a,{\cal P}^a\ra\;,\label{eq:Estate}
\end{equation}
which is a classical {\it mixed} state.\footnote{In Ref. \cite{JSTAT1}, we adopt the {\it continuous} superselection rule of $P^a$.
So, there, the {\it original} Hamiltonian is compatible with this superselection rule.}

\item[(B)]

In step (II), for example, we consider ${\cal O}$ as the compartment position in a box partitioned into left and right compartments.
Its eigenstates are the left state $|L\ra$ and the right state $|R\ra$ with their values of the wave function $\psi(L)$ and $\psi(R)$, respectively.
The process of partitioning the box is realized by increasing the height of the potential barrier in the Hamiltonian of $S_0$ between the left and right compartments in the box \cite{Szilard}.
This process is the same as the initial step of the quantum Szilard engine, which was first completely analyzed by Ref. \cite{Szilard}.

\item[(C)]

In step (III) (i.e., the non-selective measurement), the necessary condition on the apparatus $A^\prime$ is that the quantum state of $A^\prime$ has the classical fluctuation of the redefined center of mass velocity.
This condition is satisfied because, at the end of step (II), $A^\prime$ is in an energy eigenstate (\ref{eq:Estate}).

\end{enumerate}

To realize step (V) (i.e., the event reading) from step (IV), the amount of entropy production in the enlarged ensemble is required to be $\sigma=1$ nat in the combined measuring system $M$ and $-\sigma=-1$ nat in the combined measured system $S$ (note that the density matrix of the whole system is normalized) so that system $S$ is not isolated from system $M$ in their product eigenstates in the mixture (the conventional ensemble) at the end of step (IV) \cite{JSTAT1,JSTAT2}.
Otherwise, the time evolution of the state of system $S$ would {\it not} be governed by the Schr$\ddot{{\rm o}}$dinger equation in the Schr$\ddot{{\rm o}}$dinger picture \cite{JSTAT1}.

In the following, we explain this core argument.
In step (V), between two {\it state vectors} of the whole system, which are equivalent to mixtures with respect to the statistical data of the observables \cite{JSTAT2},
\begin{equation}
\sum_n|{\cal O}_n,A^\prime\rrangle_S\biggl(c_ne^{\frac{-i\delta_\mu \Xi_{A,1}}{\hbar}{\mathfrak M}_n}|{\mathfrak M}_n\ra_\psi |A_1\ra_A\biggr)_M
\end{equation}
and
\begin{equation}
\sum_n|{\cal O}_n,A^\prime\rrangle_S\biggl(c_ne^{\frac{-i\delta_\mu \Xi_{A,2}}{\hbar}{\mathfrak M}_n}|{\mathfrak M}_n\ra_\psi |A_2\ra_A\biggr)_M\;,
\end{equation}
we can trace out $A$ in $M$ to identify two {\it state vectors} of $\psi$, which are equivalent to mixtures with respect to the statistical data of the observables \cite{JSTAT2},
\begin{equation}
\sum_nc_ne^{\frac{-i\delta_\mu \Xi_{A,1}}{\hbar}{\mathfrak M}_n}|{\mathfrak M}_n\ra_\psi
\end{equation}
and
\begin{equation}
\sum_nc_ne^{\frac{-i\delta_\mu \Xi_{A,2}}{\hbar}{\mathfrak M}_n}|{\mathfrak M}_n\ra_\psi
\end{equation}
in the trigger of the event reading by $\psi$.
Here, $\delta_\mu \Xi_A=\mu\dot{\Xi}_A$ is a spatial displacement of the origin of the rearranged coordinate system $\Xi_A^a$ of $A$ in the von Neumann-type interaction\footnote{Here, we adopt the interaction picture of $M$, and we rescale time so that $\Lambda$ in Ref. \cite{EPL} is unity.} between $\psi$ and $A$ occurring over an elapsed time $\mu$ dominantly \cite{EPL}.
We can remove the dynamical degrees of freedom of $M$ from the whole system without changing the state of $S$ at the end of step (IV) if $S$ is isolated from $M$ after the inverse {\it unitary transformation} of step (IV) of the whole system (as a canonical transformation in quantum mechanics) \cite{JSTAT1,JSTAT2}; note that this unitary transformation does not change the state of $S$.
In this case, time evolution of the state of $S$ in step (V) is given by (for its derivation, see Appendix B)
\begin{equation}
\sum_nc_n|{\cal O}_n,A^\prime\rrangle_S\;,\ \{\widehat{{\cal O}}_S\}\to |{\cal O}_{n_0},A^\prime\rrangle_S\;,\ \{\widehat{{\cal O}}_S\}\;.\label{eq:Unitary}
\end{equation}
Obviously, this time evolution (i.e., the event reading process) of the state of $S$ is {\it not} governed by the Schr$\ddot{{\rm o}}$dinger equation in the Schr$\ddot{{\rm o}}$dinger picture.

Here, the entropy production in each of systems $S$ and $M$ gives rise to a non-unity normalization factor (i.e., $e^{\sigma}$ and $e^{-\sigma}$ for $S$ and $M$, respectively) in the density matrix of each system.
In the ensemble interpretation of quantum mechanics \cite{dEspagnat}, this factor redefines the pair comprising the Hilbert space of the state vectors and the space of the observables of the system \cite{JSTAT1}.
Specifically, the observables of the system are redefined by multiplying them by the factors $e^{-\sigma}$ and $e^{\sigma}$ for $S$ and $M$, respectively \cite{JSTAT1}: for its derivation, see Appendix A.2.

This redefinition is done in {\it both} the conventional and enlarged ensembles.
This simply means that {\it both} the conventional and enlarged ensembles obey the von Neumann equation with the same Hamiltonian operator except for the temporally contracted non-selective measurement process.

We add two remarks:
\begin{enumerate}
\item[(D)]
The above mixtures are with respect to the statistical data of the observables of the respective systems \cite{JSTAT2} and are described by {\it state vectors} because of the {\it reversibility} of the non-selective measurement process, which is a Hamiltonian process.
In contrast to these, the mixtures of $A$ and $A^\prime$ cannot be described by state vectors because {\it information is lost} by the orbital superselection rules (\ref{eq:SSR}) of $A$ and $A^\prime$.

\item[(E)]
The time evolution (\ref{eq:Unitary}) of the state of system $S$ is allowed in two cases: first, the state to be measured is an eigenstate; and second, system $S$ is identical to system $M$ \cite{JSTAT1}.
Thus, in both cases, the process is trace preserving in each of systems $S$ and $M$.
Specifically, the amount of entropy production in the enlarged ensemble is $0$ nat in both the systems $S$ and $M$.
\end{enumerate}

As a result of these processes, the quantum Jarzynski equality for system $S$ in the conventional ensemble is modified in two aspects \cite{JSTAT1}.
The first is the event reading process in the Heisenberg picture; in step (III), the non-selective measurement in the Heisenberg picture is incorporated in the redefinition of the Hamiltonian of system $S$ with respect to $A^\prime$ due to the macroscopicity of $A^\prime$.
Averaged over the measurement outcomes, this modification does not consequently change the original quantum Jarzynski equality \cite{MT,JSTAT1}.
The second is the entropy production in system $S$.
This modification means redefining the pair comprising the Hilbert space of the state vectors and the space of the observables of system $S$ \cite{JSTAT1}.
However, only the redefinition of the observables of system $S$ is applied in this context: see remarks (F) and (G).
As shown below, the quantum force protocol (i.e., the time dependence of the quantum part of the Hamiltonian) of system $S$ is not modified from the original quantum force protocol, which is specified by the time-dependent process from (I) to (V).
Here, because step (IV) does not change the state of system $S$, the original quantum force protocol of system $S$ does not contain the interaction Hamiltonian between system $S$ and system $M$ \cite{JSTAT1}.

The original quantum Jarzynski equality without event reading is
\begin{eqnarray}
\overline{\exp(-\beta W)}={\rm tr}_S\Biggl[\lim_{N\to \infty}{\cal T}_>\Biggl\{\prod_{n=0}^{N-1}e^{-\beta (\widehat{{\cal H}}_{H,\lambda_{t_{n+1}}}^S-\widehat{{\cal H}}_{H,\lambda_{t_n}}^S)}\Biggr\}\frac{e^{-\beta \widehat{{\cal H}}_{\lambda_0}^S}}{Z_{\lambda_0}}\Biggr]\label{eq:Jar}
\end{eqnarray}
for the variable $W$ of an amount of single-shot work\footnote{Single-shot work is deterministic and has no fluctuations.} on $S$ \cite{Talkner,Tasaki} and the controlled redefined Hamiltonian $\widehat{{\cal H}}^S_{H,\lambda_t}$ of $S$ in the Heisenberg picture with a control parameter $\lambda_t$, which is further parameterized by time $t$ in $[0,t_f]$ according to the quantum force protocol \cite{Talkner}.
Here, ${\cal T}_>$ is the time ordering of products and we set $\beta=1/(k_BT)$, $Z_\lambda={\rm tr}_S\bigl[e^{-\beta \widehat{{\cal H}}_{H,\lambda}^S}\bigr]$, and $t_n=n t_f/N$.
This equality is
\begin{eqnarray}
\overline{\exp(-\beta W)}={\rm tr}_S\Biggl[\frac{e^{-\beta \widehat{{\cal H}}_{H,\lambda_{t_f}}^S}}{Z_{\lambda_0}}\Biggr]=: \exp(-\beta \Delta F_{\rm eq})
\end{eqnarray}
and leads to the thermodynamic inequality
\begin{eqnarray}
\overline{W}&=&{\rm tr}_S\left[\widehat{{\cal H}}_{H,\lambda_{t_f}}^S\widehat{\varrho}_S(0)\right]-{\rm tr}_S\left[\widehat{{\cal H}}^S_{\lambda_0}\widehat{\varrho}_S(0)\right]\\
&\ge &\Delta F_{\rm eq}
\end{eqnarray}
by using the Jensen inequality $\overline{\exp(x)}\ge \exp(\overline{x})$ for $x=-\beta W$ \cite{Tasaki}.
Using the Jensen inequality, the modified quantum Jarzynski equality
\begin{eqnarray}
\overline{\exp (-\beta {\cal W}+3\sigma)}=\exp(-\beta \Delta F_{\rm eq})\label{eq:JarMod}
\end{eqnarray}
for the factor $e^{3\sigma}$ counter to the redefinition factor $e^{-3\sigma}$ in $\overline{\exp(-\beta {\cal W})}$ (see remarks (F) and (G)) leads to the following thermodynamic inequality:
\begin{equation}
\overline{{\cal W}}\ge \Delta F_{\rm eq}+3k_BT\;.
\end{equation}
Thus, the extra amount in the average work $\overline{{\cal W}}$ on system $S$ compared with $\overline{W}$ is $3k_BT$.
Here, in the average work $\overline{{\cal W}}$, an extra amount $k_BT$ is identified as $1$ nat entropy production $\sigma$.
This is because three redefinition factors $e^{-\sigma}$ in $\overline{\exp(-\beta {\cal W})}$ are each individually treated as a c-number constant.
Since there are two $1$ nat entropy productions in two energy event readings by the experimenter and one $1$ nat entropy production in one ${\cal O}$ event reading (i.e., one ${\mathfrak M}$ event reading) by system $M$, the experimenter and system $M$ do classical single-shot work $2k_BT$ and classical single-shot work $k_BT$ to system $S$, respectively.
Because the entropy production $\sigma$ is independent from the change of the internal energy of the measuring system, it is not accompanied by the absorption of heat \cite{JSTAT1}.
This extra amount $3k_BT$ in the average work $\overline{{\cal W}}$ compared with $\overline{W}$ can be used to experimentally test our two mutually independent theoretical results of projective quantum measurement simultaneously.
Here, the {\it average work} is the energy difference, in the quantum mechanical expectation value, of system $S$ between the initial and final times of the time-dependent process from (I) to (V).
This accords with the first law of quantum thermodynamics in the absence of heat absorption \cite{PRX}.

Here, we add two further remarks based on Appendix A.2 \cite{JSTAT1}:
\begin{enumerate}
\item[(F)] The operator to be averaged in the quantum Jarzynski equality (\ref{eq:Jar}) is defined over $t_f\ge t\ge t^{(0)}$ at an arbitrary intermediate time $t^{(0)}$ in the whole process $[0,t_f]$.\footnote{For its derivation, see Appendix C.1 of Ref. \cite{JSTAT1}.}
Thus, the operator to be averaged is redefined by multiplying it by the factor $e^{-3\sigma}$.

\item[(G)] Over the whole process $t_f\ge t^{(0)}\ge 0$, the hypothetical density matrix (the conventional ensemble) of system $S$ in the modified quantum Jarzynski equality (\ref{eq:JarMod}) is not redefined by the normalization factor $e^{\sigma}$; instead, time to define it is changed from $0$ to $t^{(0)}$.
This is because, in the average of the operator, the Hilbert space of system $S$ is fixed throughout the whole process.
\end{enumerate}

Finally, we clarify the classical nature of the single-shot work $W_{\rm e.r.}$ in the quantum mechanical context.

In quantum thermodynamics, the {\it quantum work} done by a work reservoir to a thermodynamic quantum mechanical system\footnote{In modern thermodynamics, the {\it thermodynamic systems} and the {\it systems in the thermodynamic limit} are distinct classes of systems: a microscopic system in thermal contact with a large heat bath is a thermodynamic system.} and the {\it heat absorption} from a heat bath by the system are distinguished as the change of the energy eigenvalues (the quantum Hamiltonian) of the system and the change of the corresponding occupation probabilities of the system, respectively, in the internal energy of the system \cite{Szilard,Kieu,EM}.

The discrete energy eigenvalues $\{E_n\}$ of system $S$ can be ordered as
\begin{equation}
E_0=0\;,\ E_m\ge E_{m-1}\;,\ m=1,\ldots,N\;,\ \ N\in{\mathbb{N}}\;.
\end{equation}
After the classical single-shot work $3W_{\rm e.r.}$ is done to system $S$, this structure changes to $\left\{E^\prime_{n,\{\sigma_i\}}\right\}$ as
\begin{equation}
E^\prime_{n,\{\sigma_i\}}=E_n+ \sum_{i=1}^3k_BT \sigma_i\;,\ n=0,1,\ldots,N\;,\ \sigma_i=0,1\;.\label{eq:CSS}
\end{equation}
Here, the reason why the classical part $\{\sigma_i\}$ of the energy eigenvalues $E^\prime_{n,\{\sigma_i\}}$ (i.e., the classical part of the Hamiltonian) is discrete is that the entropy production $\sigma=1$ is partitioned entirely to system $M$.

\begin{figure}[htbp]
\begin{center}
\includegraphics[width=0.6 \hsize, bb=1 4 260 231]{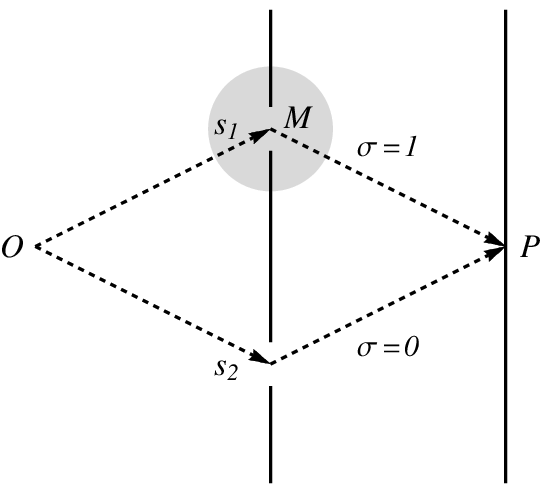}
\end{center}
\caption{A double-slit-type experiment for system $S$ to measure the amount of $W_{\rm e.r.}$ after the non-selective measurement in system $S$.
Besides the screen, there is the source point (or the entrance slit) $O$ of system $S_0$ and there are two slits $s_1$ and $s_2$, which select the paths of system $S_0$.
System $M$ (the gray zone) is located at slit $s_1$ and interacts with system $S$ there only.
Two wave functions of system $S$ with $\sigma=1$ and $\sigma=0$ interfere with each other at each point $P$ on the screen through slits $s_1$ and $s_2$, respectively.}
\end{figure}

Equation (\ref{eq:CSS}) shows the classical nature of the single-shot work $W_{\rm e.r.}$ in the quantum mechanical context.
Even with the classicality of $W_{\rm e.r.}$, the amount of $W_{\rm e.r.}$ is measurable.
To measure the amount of $W_{\rm e.r.}$, we consider a double-slit-type experiment for system $S$ after the non-selective measurement is done in system $S$ (see Fig. 2).
We denote the two slits by $s_1$ and $s_2$.
We assume that an event reading of system $S$ is induced by another system $M$ at slit $s_1$ so that the classical states of system $S$ at slits $s_1$ and $s_2$ are the excited state (i.e., $\sigma=1$) and the ground state (i.e., $\sigma=0$), respectively.
Then, the temporal interference pattern, on the screen, of the two wave functions of system $S$ determines the amount of $W_{\rm e.r.}$.
Here, the classical energy-eigenvalue difference gives rise to the angular frequency difference between the two classical energy eigenfunctions.

\section{Conclusion}

In this paper, first, we refined the derivation of the result (\ref{eq:0}) of Ref. \cite{JSTAT1} in quantum mechanics totally governed by the Schr${\ddot{\rm{o}}}$dinger equation in the Schr$\ddot{{\rm o}}$dinger picture.
Second, we combined this result and the result of Ref. \cite{EPL} to model a Hamiltonian process in the setting of the quantum Jarzynski equality involving four specified systems (\ref{eq:6}), which can be used to test these two mutually independent theoretical results of projective quantum measurement simultaneously.

Here, the previous result (\ref{eq:0}) can be intuitively understood as follows.
Just before the event reading, the combined measuring system $M$ loses 1 nat of information and the combined measured system $S$ acquires this 1 nat of information.
Then, system $S$ is not isolated from system $M$ in their product eigenstates in a mixture at this time \cite{JSTAT1,JSTAT2}.
To cancel this non-isolation of $S$ from $M$, the amount of classical single-shot work that $M$ is required to apply to $S$ is $k_BT$ in the setting of the quantum Jarzynski equality with respect to $S$.
This situation is oppositely analogous to (i.e., the work has the opposite sign but the mechanism is analogous) the measurement-based feedback process by which an agent (e.g., Maxwell's demon), who acquires the QC-mutual information amounting to $I_{\rm QC}$ nats by a POVM measurement of a system, can extract quantum work amounting to $k_BTI_{\rm QC}$ \cite{SU1,MT,Nature1}.

Finally, to conclude this paper, we explain how this result can be positioned within the existing literature on quantum thermodynamics from the perspective of the information-theoretical origins of work, in the setting of the quantum Jarzynski equality, in terms of the density matrix.
The explanation consists of three stages.

The first stage is the second law of quantum thermodynamics without incorporating quantum measurements in the analysis of the work.
This law states that
\begin{equation}
\overline{W}\ge \Delta F\label{eq:WF1}
\end{equation}
for the average quantum work $\overline{W}$ done by a work reservoir to a thermodynamic quantum mechanical system, which is in thermal contact with a large heat bath at temperature $T$, and the change in the non-equilibrium free energy $\Delta F$ of the system\footnote{Non-equilibrium free energy is obtained from the equilibrium free energy by replacing the thermodynamic entropy with information-theoretical entropy \cite{Sagawa}.} induced by the work in $\overline{W}$.
In general, the work $W$ fluctuates thermally and quantum mechanically.
In the thermodynamic limit of the system, the thermal fluctuations in the system become negligible.
When the change of the Hamiltonian of the system is cyclic, the change in the non-equilibrium free energy $\Delta F$ of the system has an information-theoretical origin in terms of the density matrix:
\begin{equation}
\Delta F=k_BTS(\widehat{\varrho}|| \widehat{\varrho}_{\rm can})\;,\label{eq:WF2}
\end{equation}
where $\widehat{\varrho}$ is the terminal quantum state of the system and $\widehat{\varrho}_{\rm can}$ is the initially prepared canonical thermal equilibrium quantum state of the system at temperature $T$ \cite{Donald}.
Here, the relative von Neumann entropy is always non-negative: $\Delta F\ge 0$.
It should be noted that this information-theoretical expression (\ref{eq:WF1}) and (\ref{eq:WF2}) for the work was extended, in the single-shot work, to the Gibbs-preserving and completely-positive and trace-preserving map of the system in the context of quantum resource theory, where the extractable work from a resource and the work required to create a resource are each individually regarded as the work and the thermodynamic limit of the system is unnecessary \cite{Sagawa,Res1,Res2,Res3,ResReview}.

The second stage is to incorporate quantum measurement, the measurement-based feedback process, and the reset process of the measurement outcome stored in the memory system (i.e., the agent) in the context of the quantum thermodynamics of information \cite{SU1,Nature1,SU2}.
In this theory, quantum measurement is treated as a projective quantum measurement that is done in the ensemble description \cite{SU1,MT} and in the general POVM measurement, which is the projective quantum measurement of the memory system (not of the composite system) after entangling the system with the memory system by a unitary time evolution of the composite system, and the physical substance of event reading (i.e., the projection hypothesis) is treated as a black box and is not considered.
As mentioned above, the extractable quantum work $-W_{\rm f.b.}$ by the measurement-based feedback process has the information-theoretical origin
\begin{equation}
-W_{\rm f.b.}\le k_BTI_{\rm QC}\;,\label{eq:Wfb1}
\end{equation}
where the QC-mutual information $I_{\rm QC}$ satisfies $0\le I_{\rm QC}\le H(\{p\})$ for the Shannon entropy $H(\{p\})$ of the outcome probabilities $\{p\}$ in the POVM measurement and reduces to the classical mutual information in the case of a classical measurement \cite{SU1}.
However, to perform projective quantum measurement in the ensemble description \cite{SU1,MT} and to reset the memory, quantum work $W_{\rm meas}$ and quantum work $W_{\rm reset}$ are required to be done to the thermodynamic memory system, respectively.
The lower bound of their sum is given by \cite{SU2}
\begin{equation}
W_{\rm meas}+W_{\rm reset}\ge k_BT I_{\rm QC}\;,\label{eq:Wfb2}
\end{equation}
which is the comprehensive form of the Landauer principle for the work required to reset the memory \cite{Landauer1,Landauer2}.
The upper bounds of the extractable quantum work in Eqs. (\ref{eq:Wfb1}) and (\ref{eq:Wfb2}) cancel each other out.
Note that there have been recent advances in the measurement-based quantum heat engine: for a review, see Ref. \cite{AVS}.
The key novel steps in this type of quantum heat engine are projective quantum measurement, the measurement-based feedback process, and the memory reset process, all of which are fundamentally described here.

The third stage is to incorporate the work required for event reading, that is, the informatical process in projective quantum measurement $W_{\rm e.r.}$, which is independent from quantum work $W_{\rm meas}$.
As noted, this work required for event reading was first analyzed by Refs. \cite{JSTAT1} and \cite{JSTAT2}.
In the notation of Appendix A.2, the information-theoretical origin of this classical single-shot work in terms of the density matrix is
\begin{eqnarray}
W_{\rm e.r.}&=&k_BTS(\widehat{\varrho}_M || \widehat{\varrho}_M^\star)\\
&=&-k_BTS(\widehat{\varrho}_S || \widehat{\varrho}_S^\star)\;,
\end{eqnarray}
where $S(\widehat{\varrho}||\widehat{\varrho}^\star)$ is the generalized relative von Neumann entropy:
\begin{eqnarray}
S(\widehat{\varrho}_M || \widehat{\varrho}_M^\star)&=&1\;,\\
S(\widehat{\varrho}_S || \widehat{\varrho}_S^\star)&=&-1\;.
\end{eqnarray}
Here, $W_{\rm e.r.}$ is internal work in the composite system of systems $S$ and $M$.
Namely, in the composite system, the work for $S$ and that for $M$ cancel each other out and there is no net extractable work and no net energy cost.

We now conclude this paper.
As seen from the above explanations of the information-theoretical origins of work in quantum thermodynamics in three stages, the extractable quantum work $-W_{\rm f.b.}$ characterizes the information-to-energy conversion (see Ref. \cite{Nature2} for a renowned experiment in the classical mechanical context), but the classical single-shot work $W_{\rm e.r.}$ characterizes the physicality of the projection hypothesis in the setting of the quantum Jarzynski equality.
Namely, in the ensemble interpretation of quantum mechanics, we have theoretically shown a way to experimentally examine the physicality of the projection hypothesis \cite{JSTAT2}.
This is the implication of the main results in the fields of quantum information and quantum thermodynamics.

\begin{appendix}

\setcounter{equation}{0}
\renewcommand{\theequation}{\Alph{section}.\arabic{equation}}

\section{Density Matrix Descriptions}

In this appendix, we present supplementary calculations for the density matrix descriptions.

\subsection{Original von Neumann equations for Eqs. (\ref{eq:1}) and (\ref{eq:2})}

In this subsection, we present the original von Neumann equations of a density matrix for Eqs. (\ref{eq:1}) and (\ref{eq:2}) before truncation of the temporally contracted non-selective measurement process under the identification of the statistical weight with the normalization factor of the density matrix.

In the direct description, the original von Neumann equation of the density matrix, $\widehat{\varrho}(t)$, is
\begin{equation}
\widehat{\varrho}(dt)=\sum_{{\rm all}\ y}\widehat{P}(y)\widehat{\varrho}(0)\widehat{P}(y)\;,
\end{equation}
where $\widehat{P}(y)$ is the projection operator of a measurement outcome $y$.
After truncation of the temporally contracted non-selective measurement process (i.e., replacement of the right-hand side with zero), this equation reduces to Eq. (\ref{eq:1}).

In the statistical description, the original von Neumann equation of the density matrix, $\widehat{\rho}(t)$, is
\begin{equation}
\widehat{\rho}(dt)=(1-\delta(t)dt)\widehat{\rho}(0)+\delta(t)dt\sum_{{\rm all}\ y}\widehat{P}(y)\widehat{\rho}(0)\widehat{P}(y)\;.\label{eq:original2}
\end{equation}
After truncation of the temporally contracted non-selective measurement process (i.e., replacement of the second term on the right-hand side with zero), this equation reduces to Eq. (\ref{eq:2}).

Note that this truncation of the temporally contracted non-selective measurement process is not a trace-preserving map of the density matrix $\widehat{\rho}(t)$.
However, this violation of trace preservation is attributable to the c-number normalization factor $e^{-1}$ that arises after the truncation.
Thus, in the statistical description, the process after the truncation is compatible with the original quantum mechanical process (\ref{eq:original2}) before the truncation.

\subsection{Redefinitions of systems by the entropy production (\ref{eq:4})}

In this subsection, we derive the redefinitions of systems $S$ and $M$ by the entropy production (\ref{eq:4}) \cite{JSTAT1}.

For the entropy production $\sigma_X$ in each of systems $X$ ($X=S$, $M$), the density matrix $\widehat{\varrho}_X$ (the conventional ensemble) of system $X$ is redefined by $\widehat{\varrho}_X^\star$ as
\begin{equation}
\widehat{\varrho}_X^\star=e^{-\sigma_X}\widehat{\varrho}_X\;.\label{eq:varrho}
\end{equation}

Then, the entropy production $\sigma_X$ can be regarded as the generalized relative von Neumann entropy
\begin{eqnarray}
\sigma_X&=&S\left(\widehat{\varrho}_X || \widehat{\varrho}_X^\star \right)\\
&:=&{\rm tr}_X\left[\widehat{\varrho}_X \ln \widehat{\varrho}_X-\widehat{\varrho}_X \ln \widehat{\varrho}_X^\star\right]\;.
\end{eqnarray}
Here, the relative von Neumann entropy is generalized to allow in the second slot the redefinition of the Hilbert space of the state vectors by the entropy production, and then it does not satisfy the non-negativity property.

Simultaneously with Eq. (\ref{eq:varrho}), we consider the following observables $\widehat{{\cal O}}_X^\star$ of system $X$ redefined from the original observables $\widehat{{\cal O}}_X$ of system $X$ as
\begin{equation}
\widehat{{\cal O}}_X^\star=e^{\sigma_X}\widehat{{\cal O}}_X\;.
\end{equation}
For $\widehat{{\cal O}}_X^\star$, the density matrix $\widehat{\varrho}_X^\star$ gives their original expectation values by the next trace operation:
\begin{equation}
{\rm tr}_X\left[\widehat{{\cal O}}_X^\star \widehat{\varrho}_X^\star\right]={\rm tr}_X\left[\widehat{{\cal O}}_X\widehat{\varrho}_X\right]\;.
\end{equation}
From this fact, for $\widehat{{\cal O}}_X^\star$, the density matrix $\widehat{\varrho}_X^\star$ is regarded as the original ensemble of system $X$ after the redefinition.

In Sec. 4, we derived
\begin{equation}
\sigma_S=-\sigma_M=-1
\end{equation}
under the condition on the time evolution of system $S$ summarized in Appendix B.

\setcounter{equation}{0}
\renewcommand{\theequation}{\Alph{section}.\arabic{equation}}

\section{Unitary Transformation (\ref{eq:Unitary})}

In this appendix, we explicitly write out the unitary transformations that would result in the time evolution (\ref{eq:Unitary}) of system $S$.
In the following, we assume that $S$ was isolated from $M$ after the unitary transformations.
\begin{enumerate}
\item[(a)]
Before step (IV), systems $S$ and $M$ are decoupled from each other:
\begin{equation}
\left\{\left(\begin{array}{c} |S_n\rrangle \\ \\ \{\widehat{{\cal O}}_S\}\end{array}\right)\right\}_n\;,\ \left(\begin{array}{c} |M_0\rrangle \\ \\ \{\widehat{{\cal O}}_M\}\end{array}\right)\;.
\end{equation}
In this notation, the {\it state vector} of $S$ is equivalent to a mixture with respect to the statistical data of the observables of $S$ \cite{JSTAT2}.

\item[(b)]
After step (IV), systems $S$ and $M$ are entangled with each other.
However, after the unitary transformation (u.t.), their state vectors can be decoupled from each other:
\begin{eqnarray}
\left\{\left(\begin{array}{c} |S_n\rrangle \\ \\ \{\widehat{{\cal O}}_S\}\end{array}\right)\;,\ \left(\begin{array}{c} |M_n\rrangle \\ \\ \{\widehat{{\cal O}}_M\}\end{array}\right)\right\}_n
&\overset{{\rm u.t.}}{\Rightarrow}&
\left\{\left(\begin{array}{c} |S_n\rrangle \\ \\ \{\widehat{{\cal O}}_S\} \end{array}\right)\;,\ \left(\begin{array}{c}|M_0\rrangle \\ \\ \{\widehat{U}_{\rm IV}^{(n)\ -1}\widehat{{\cal O}}_M\widehat{U}_{\rm IV}^{(n)}\}\end{array}\right)\right\}_n\nonumber\\
&\overset{{\rm tr}_M}{\Rightarrow}&
\left\{\left(\begin{array}{c} |S_n\rrangle \\ \\ \{\widehat{{\cal O}}_S\}\end{array}\right)\right\}_n\;,
\end{eqnarray}
where ${\rm tr}_M$ denotes the removal of the dynamical degrees of freedom of $M$ from the whole system.
This is because step (IV) does not change the state of $S$.

\item[(c)] 

After step (V), the same argument as (b) holds:
\begin{eqnarray}
\left\{\left(\begin{array}{c} |S_n\rrangle \\ \\ \{\widehat{{\cal O}}_S\}
\end{array}\right)\;,\ \left(\begin{array}{c} \widehat{U}_{{\rm V}}|M_n\rrangle \\ \\ \{\widehat{{\cal O}}_M\}\end{array}\right)\right\}_n
&\overset{{\rm u.t.}}{\Rightarrow}&
\left\{\left(\begin{array}{c} |S_n\rrangle \\ \\ \{\widehat{{\cal O}}_S\} \end{array}\right)\;,\ \left(\begin{array}{c}|M_0\rrangle \\ \\ \{\widehat{U}_{{\rm IV}}^{(n)\ -1}\widehat{U}_{{\rm V}}^{-1}\widehat{{\cal O}}_M\widehat{U}_{{\rm V}}\widehat{U}_{{\rm IV}}^{(n)}\}\end{array}\right)\right\}_n\nonumber\\
&\overset{{\rm tr}_M}{\Rightarrow}&
\left\{\left(\begin{array}{c} |S_n\rrangle \\ \\ \{\widehat{{\cal O}}_S\}\end{array}\right)\right\}_n\;.
\end{eqnarray}

\item[(d)] 
After the event reading, the states of systems $S$ and $M$ become the following:
\begin{eqnarray}
\left(\begin{array}{c} |S_{n_0}\rrangle \\ \\ \{\widehat{{\cal O}}_S\}\end{array}\right)\;,\ \left(\begin{array}{c} |M_{n_0}\rrangle \\ \\ \{\widehat{{\cal O}}_M\}\end{array}\right)\overset{{\rm tr}_M}{\Rightarrow}\left(\begin{array}{c} |S_{n_0}\rrangle \\ \\ \{\widehat{{\cal O}}_S\}\end{array}\right)\;.
\end{eqnarray}
\end{enumerate}

\setcounter{figure}{0} 
\renewcommand{\thefigure}{\Alph{section}.\arabic{figure}}

\section{Experimental Platform}

In this appendix, we clarify the experimental platform for realizing the four systems which appear in the quantum thermodynamic scheme proposed in Sec. 4.
A model of the quantum measurement steps in the proposed scheme is shown in Fig. C.1.

\begin{figure}[htbp]
\begin{center}
\includegraphics[width=0.4 \hsize, bb=0 0 260 232]{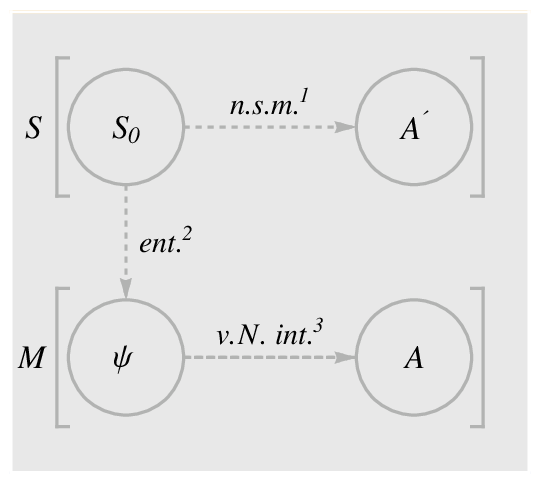}
\end{center}
\caption{Quantum measurement steps in the proposed quantum thermodynamic scheme.
There are three steps: first, the non-selective measurement (n.s.m.) in the composite system $S=S_0+A^\prime$; second, the entangling (ent.) process between systems $S_0$ in $S$ and $\psi$ in $M$; and third, the von Neumann-type interaction (v.N. int.) between systems $\psi$ and $A$.
Each dashed arrow from system $X$ to system $Y$ indicates the interaction process (e.g., the von Neumann-type interaction) between systems $X$ and $Y$, where the state of system $X$ does not change.}
\end{figure}

In Fig. C.1, $S_0$ and $\psi$ are quantum mechanical measured and event-reading systems, respectively.
In the context of quantum measurement theory \cite{dEspagnat}, they are distinguished from each other.
See also remark (E).

A macroscopic apparatus $A^\prime$ is the separation apparatus (i.e., a macroscopic detector: see Refs. \cite{Araki1,NM1,NM2} for examples of macroscopic detectors) in the non-selective measurement.
The meaning of the {\it separation apparatus} is that it performs the non-selective measurement in the composite measured system $S$ in the presence of the orbital superselection rule of this apparatus by separating the superselection sectors of system $S$, but the quantum state of the separation apparatus cannot record the non-selective measurement results \cite{JSTAT1,Araki1,Araki2}.

In the context of quantum measurement theory, the novel element in the four systems is system $A$.
It is a macroscopic BEC {\it in quantum field theory}, which breaks the spatial translational symmetry spontaneously, in the presence of the orbital superselection rule.
As such the BEC, the photon condensate in the equilibrium superradiant phase transition (SPT) (i.e., the spontaneous creation of non-zero and static classical electromagnetic field and electromagnetic polarization at a thermal equilibrium) in cavity quantum electrodynamics (QED), where we treat the cavity wall (i.e., the boundary condition) as part of this system, is a desirable system \cite{EPL}.

To date, there is a five-decade history of theoretical studies of the SPT since it was first predicted by Hepp and Lieb \cite{HL1,HL2} in the ultrastrongly coupled Dicke model; here, the Dicke model of two-level atoms and radiation \cite{Dicke} is a benchmark model of the cavity QED.
However, the no-go theorem for the SPT has been shown when we consider the gauge-invariant model \cite{NG1,NG2}.
Here, note that the Dicke model truncates the diamagnetic term from the Hamiltonian artificially and thus loses its gauge invariance.
However, following the refined no-go theorems for the SPT proven in Refs. \cite{NG3} and \cite{NG4}, three theoretical articles have been published \cite{SQPT1,SQPT2,SQPT3}.
These articles evade the no-go theorem for the SPT in the gauge-invariant cavity QED model by removing the basic assumption of spatial uniformity of the quantum electromagnetic field within the cavity: the quantum electromagnetic field spatially varies within the cavity.
Here, the SPT results from a magnetostatic instability \cite{SQPT3}.
Though these theoretical predictions for the SPT have not yet been experimentally confirmed, such an SPT could be used as a platform for realizing system $A$ as a macroscopic BEC in quantum field theory and experimentally testing the predictions of our quantum thermodynamic scheme proposed in Sec. 4, that is, realization of the projection hypothesis in projective quantum measurement and the work required for an event reading.

\end{appendix}

\end{document}